\documentclass[aps,twocolumn,showpacs,preprintnumbers]{revtex4}
\usepackage{graphicx}

\begin{document}

\title{On recombination in strong laser fields: effect of a slow drift}

\author{Jacek Matulewski}%
\email{jacek@phys.uni.torun.pl}
\author{Andrzej Raczy\'nski}%
\author{Jaros\l aw Zaremba}

\affiliation{Instytut Fizyki, Uniwersytet Miko\l aja Kopernika,\\
ul. Grudzi\c{a}dzka 5, 87-100 Toru\'n, Poland}%

\pacs{34.50.Rk, 32.80.Wr, 34.80.Lx}


\begin{abstract}
The  dynamics of the recombination in ultrastrong atomic fields is
studied for one-dimensional models by numerical simulations. A
nonmonotonic behavior of the bound state final population as a
function of the laser field amplitude is examined. An important
role of a slow drift of an electron wave packet is observed.
\end{abstract}

\maketitle

\section{Introduction}

It is well known that during atomic photoionization in ultrastrong
laser fields a peculiar effect occurs: for a given pulse length an
interval of field intensities exists in which the ionization
probability is a decreasing  function of the field intensity. This
is called adiabatic atomic stabilization against photoionization
\cite{gavrila,fedorov}. The motivation of the present work is to
look for a similar effect in the inverse process, namely atomic
recombination in ultrastrong fields. The main difference between
these processes is due to different initial conditions. The
process of strong-field recombination has recently been examined
by Hu and Collins \cite{hu}, who however investigated the
dependence of the process on the initial momentum of the incoming
electron.

During the recombination one should expect dynamical effects
similar to those observed in the case of ionization. As it has
been shown in the works devoted to the atomic stabilization, in
laser fields of order of a few atomic units most of the electron
wave packet moves as a whole, performing oscillations in the
rhythm of the field. Additionally, long time oscillations of the
packet are possible which are due to an asymmetry of the
interaction of its different parts with the binding potential; in
our earlier papers \cite{mrz-stab,mrz-mg1,mrz-mg2} we have called
them a slow drift. The slow drift has the range equal to that of
the classical oscillations of a free electron in the laser field
and it was possible to give an analytic formula for its frequency
in the case of the binding potential being a rectangular well. The
fact that a significant part of the packet moves without changing
of its shape, in particular of its width, can be explained in
terms of the so-called Kramers-Henneberger (KH) well, which is the
mean potential (zero-th term of the Fourier expansion) of the
oscillating nucleus in the electron's reference frame. The
stabilization of the wavepacket can be interpreted as its trapping
in the eigenstate of the KH well. In the laboratory frame the KH
well oscillates in the rhythm of the laser field and may also
perform the slow drift, which was described by a generalized model
of Ref. \cite{mrz-stab}. For longer times a decay of the KH state
can be observed, due to an influence of higher terms of the
above-mentioned Fourier expansion, and the corresponding decay
rate has been evaluated \cite{vps}. The probability of finding the
electron in the atomic bound state is thus a complicated function
of time, which reflects a rapid oscillatory motion, a slow drift
and a finite lifetime of the KH eigenstate. Moreover, a necessary
condition for the stabilization to occur is that the pulse have
such a shape that the corresponding electron classical trajectory
should remain in the neighborhood of the nucleus.

In this work we present the results of numerical simulations of
the dynamics of recombination for two model one-dimensional
systems: a rectangular potential well and a long-range
Coulomb-like potential. We will examine the time dependence of the
bound state population (initially equal to zero) and the dynamics
of the wave packet. As the measure of efficiency of recombination
we take the final population of the bound state. In particular we
study how the details of the dynamics depend on the amplitude of
the laser field and as a consequence on the details of the slow
drift.

\section{Numerical approach}
Strong-field recombination has been investigated in which an
initially free electron modeled by a Gaussian packet, intially
resting at twice the amplitude of classical free oscillations, is
carried towards the ion in the laser field. The ion has been
modeled by two kinds of a binding potential.

The dynamics of recombination has been examined by numerically
solving the time-dependent Schr\"odinger equation in one spatial
dimension. The computations have been performed on a grid of
$16384$ points. The space covered by our grid extends from $-100$
to $100$ a.u. (the grid step is about $0.0122$ a.u.). Calculations
have been performed for two atomic models: the rectangular
symmetrical well and a soft core atom potential. The rectangular
well
\begin{equation}
V(x)=-V_1 \Theta(|a_1-x|)
\end{equation}
of the depth equal to $V_1=2.049$ a.u. and half-width equal to
$a_1=0.122$ a.u. is placed in the middle of the simulation space.
The function $\Theta$ is the Heaviside step function. The same
well parameters were earlier used in the studies of stabilization
\cite{mrz-stab}. We have also used the model atom potential
proposed by Eberly {\em  et al.} \cite{es}:
\begin{equation}
V(x)=-\frac{1}{\sqrt{1+x^2}}. \label{eq-atom}
\end{equation}
To simulate the evolution of a one-dimensional wavefunction we use
the Crank-Nicholson scheme, which for Schr\"odinger equation leads
to the often used symmetrical Cayley's formula \cite{nrcpp}. The
simulations have been performed in the dipole approximation in the
length gauge, i.e. the Hamiltonian of the system reads (in atomic
units)
\begin{equation}
\hat{H}=-\frac{1}{2} \frac{\partial^2}{\partial x^2}+ V(x)+ x
\cdot \varepsilon(t),
\end{equation}
where the laser electric field $\varepsilon(t)=\varepsilon_0
\Theta(t)\Theta(t_{f}-t)\cos(\omega t-\pi)$ is directed along the
$x$-axis and acts only during the time interval $t_{f}$. The
period of the laser is $T=2\pi /\omega$, where $\omega=1$ a.u. is
the laser frequency. We have changed the laser field amplitude
$\varepsilon_0$ in the range between $1$ to $15$ a.u. The time
step was equal to $\Delta t=T/10^4$. The convergence of the
computer simulations has been checked on the grid with $65536$
spatial nodes and alternatively with a ten-times smaller time
step. No important differences have been visible.

The initial position of the wavepacket, initially taken to be
Gaussian $|\psi|^{2}=\exp[-(x-x_0)^{2}/(2\delta^{2})]$ with the
variance $\delta^{2}=0.25$ a.u., has been prepared very carefully,
independently for each simulation. The main condition is that the
electron should be placed over the well after the first half of
the period, i.e. $\bar{x}(T/2)=0$, $\bar{x}$ being the packet's
center of mass. Moreover, we require it to have at that time zero
velocity to make the recombination more probable. This is why we
have chosen the packet's initial position $x_0=-2\varepsilon_0 /
\omega^2$ and initial velocity equal to zero so that the first
classical turning point of a free oscillation occurs over the
well. Note that the dependence of the recombination efficiency on
the initial packet velocity has already been the subject of
investigations of Ref. \cite{hu}.
\\

\section{Results and interpretation}

\begin{figure}[h]
\centering
\includegraphics[width=0.7\columnwidth,angle=-90]{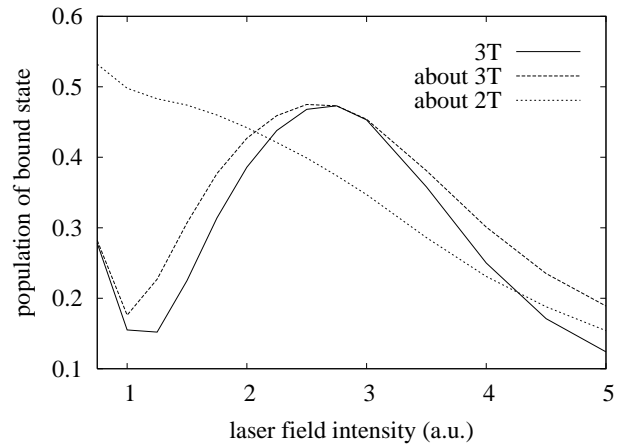}
\caption {\label{fig1} The population of the ground state in the
model atomic potential in the recombination process as a function
of the field intensity $\varepsilon_{0}$ in the case of the field
being switched off at exactly 3$T$, solid line; at the time
instant of a maximum population in the vicinity of 3$T$ (see
comment in the text), dashed line; at the time instant of a
maximum population in the vicinity of 2$T$, short-dashed line.}
\end{figure}

\begin{figure}[h]
\centering
\includegraphics[width=0.7\columnwidth,angle=-90]{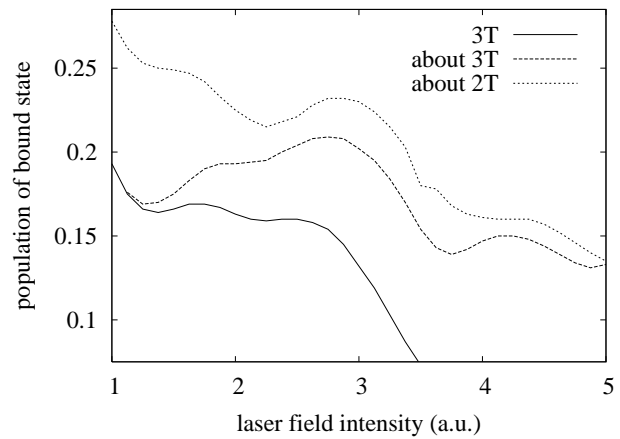}
\caption {\label{fig2} As in Fig. \ref{fig1} but for an only bound
state of a well potential}
\end{figure}

\begin{figure}[h]
\centering
\includegraphics[width=0.9\columnwidth,angle=0]{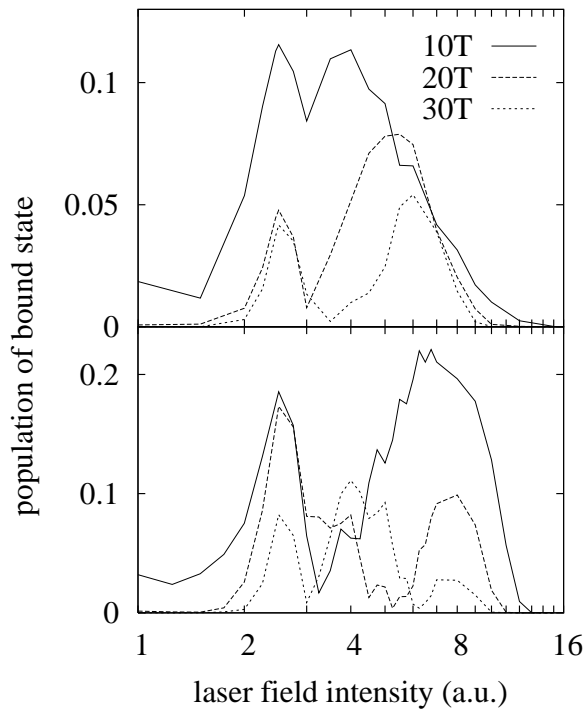}
\caption {\label{fig3} The population of the bound state of the
well potential in the recombination process (upper plot) and
ionization process (lower plot) as functions of the field
intensity $\varepsilon_{0}$ in the case of the field being
switched off at exactly 10$T$, 20$T$ and 30$T$.}
\end{figure}

In the case of the recombination process one observes a
nonmonotonic dependence of the bound state population on the field
intensity. It is well visible in Fig. \ref{fig1} where the ground
state population is shown as a function of the field intensity. In
each case the rectangular pulse has been switched off at the time
instant closest to three optical periods (3$T$) at which the
ground state population exhibits a maximum (except for the solid
line which has been obtained exactly at 3$T$). At the time 2$T$
the maximum is not yet formed. A wide maximum is seen for
$\varepsilon_{0}\approx 2.5$ a.u. both for the pulse duration of
exactly 3$T$ and approximately 3$T$. A similar nonmonotonicity is
observed in the case of a potential well (see Fig. \ref{fig2}).
The figure suggests an existence of even more maxima which are
however less pronounced for short times. This is confirmed for
longer pulses in Fig. \ref{fig3} (upper plot) in which one can see
two separated maxima, one of which occurs for
$\varepsilon_{0}\approx 2.5$ a.u. for all pulse lengths, while the
other one, being wider, shifts slowly for increasing pulse
lengths. The origin of the peaks is different. The left peak can
be attributed to a long lifetime $\tau$ of the eigenstate of the
Kramers-Henneberger well; according to Ref. \cite{vps} $\tau$
exhibits a maximum for such $\varepsilon_{0}$ for which
$J_{1}(\varepsilon_{0}\sqrt{2\omega}/\omega^{2})=0$, i.e
$\varepsilon_{0}=$2.71 a.u. and 4.96 a.u. (for the range of field
amplitudes examined in this work). The position of the right hand
side peak for different values of the field intensity and of the
pulse duration is evidently connected with the slow drift of the
wave packet. A maximum is observed if at the moment of the field
switch-off the slow drift of the main part of the packet is at
such a stage that it is located at the well, which does not
usually occur after an entire number of periods, and has a
velocity close to zero. A similar field dependence of the bound
state population can be seen in the case of the ionization (lower
plot). Again one can see a peak for $\varepsilon_{0}\approx 2.5$
a.u. for all the pulse durations and a second peak the position of
which changes. One can also notice a small peak at about 5 a.u.
(second minimum of $J_{1}$ and thus of $1/\tau$).

\begin{figure}[h]
\centering
\includegraphics[width=0.45\columnwidth,angle=-90]{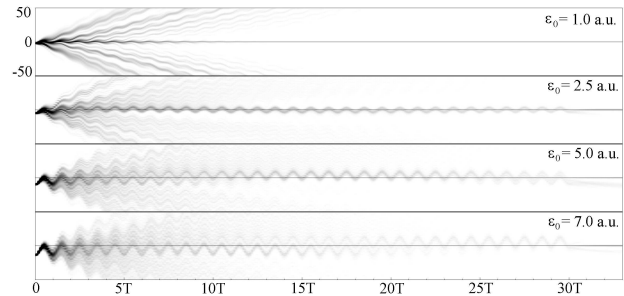}
\caption {\label{fig4} The time evolution of the wave packet
recombination in the well potential for different values of the
field intensity. The level of blackening is proportional to the
squared modules of the wave function.}
\end{figure}

\begin{figure}[h]
\centering
\includegraphics[width=0.45\columnwidth,angle=-90]{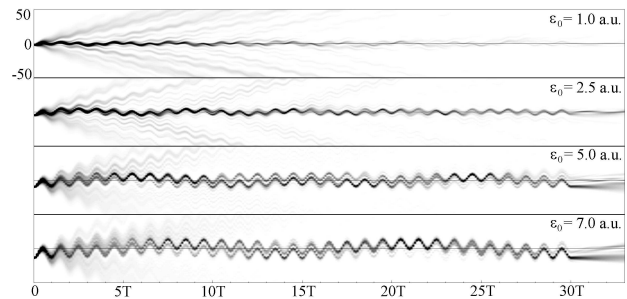}
\caption {\label{fig5} As in Fig. \ref{fig4} but for a model
atomic potential.}
\end{figure}

In Figs \ref{fig4} and \ref{fig5} we show the evolution of the
wave packet for different values of the field amplitude, for the
potential well and for the model atom, respectively. For
relatively small field amplitudes each field oscillation causes
tearing the packet into smaller parts, typical of the regime of
multiphoton ionization. For the field amplitude being of order of
2.5 a.u. one can already distinguish a main part of the packet,
represented by the oscillating streak. A slow drift is imposed on
the rapid oscillations with the field frequency $\omega$. For the
potential well the frequency of the slow oscillations may be
evaluated using the formula
$\Omega=\sqrt{2a_{1}V_{1}/(\sqrt{2\pi}\sigma^{3})}$, where
$\sigma=2\varepsilon_{0}/\omega^{2}$ is the width of the wave
packet trapped in the KH well \cite{mrz-stab}. Due to the slow
drift, after an entire number of field periods, the packet ends up
at different distances from the potential center, depending on the
field amplitude.

The influence of the atomic potential is for our data stronger
then in the case of the potential well, which is manifested by a
larger frequency of the slow drift and a more effective
recombination and population trapping (darker line) and less
population thrown away as isolated small portions. In the case of
the well potential, for large times the streak for
$\varepsilon_{0}=2.5$ a.u. is significantly darker than for 2.0
a.u. and 3.0 a.u. (not shown), which corresponds to the first
maximum of Fig. \ref{fig3} (upper plot). There is also a
correspondence between the level of population shown in Fig.
\ref{fig3} (upper plot) and the position of the streak of Fig.
\ref{fig4}. For example for $t=10T$ and $\varepsilon_{0}=5$ a.u.
the streak is located at the position of the well $x=0$ while for
$\varepsilon_{0}=7$ a.u. its distance from the well is close to
maximum due to the slow drift; this corresponds to the fact that
the bound state population in Fig. \ref{fig3} (upper plot) for
$\varepsilon_{0}=5$ a.u. is larger than for $\varepsilon_{0}=7$
a.u. For  $t=30T$ the situation is reversed due to the slow drift.

\begin{figure}[h]
\centering
\includegraphics[width=0.7\columnwidth,angle=-90]{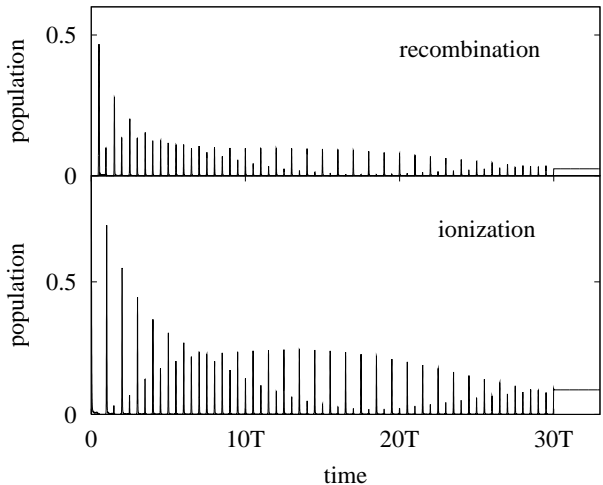}
\caption {\label{fig6} The time evolution of the bound state
population in the case of a well potential for the field intensity
$\varepsilon_{0}=5$ a.u.}
\end{figure}

For larger fields, especially in the case of atoms, the streak
splits into two ones, which corresponds to population trapping in
a superposition of two eigenstates of the KH well. One can also
see a flow of the population from one streak to the other, which
corresponds to the population oscillating between those two
states. In the case of the rectangular well, for which an
analytical form of the KH well is known \cite{mrz-stab}, we have
verified that for the field amplitude $\varepsilon_{0}=$ 7 a.u.
there are indeed two eigenstates, the energy difference of which
is approximately 0.016 a.u.; the corresponding period of
oscillations agrees with the period of the population flow of Fig.
\ref{fig4}.

The presence of the slow drift is reflected in the time dependence
of the bound state population visible in Fig. \ref{fig6}. The
maxima of the comb-like structure appear twice in each optical
cycle at the time instants at which the packet's velocity is zero
(turning points). The heights of the maxima depend on the current
stage of the slow drift. The frequency of the exchange of the two
families of the peaks, appearing respectively after an even or odd
number of half-cycles, is the same, both in the case of ionization
and recombination, as the frequency $\Omega$ of the slow drift,
visible also in Fig. \ref{fig4}.
\\

\section{Conclusions}
A nonmonotonic behavior of the bound state final population as a
function of the field amplitude has been observed in numerical
simulations of strong-field recombination for one-dimensional
models. The maxima may have two origins. First, they appear for
such laser field intensities and pulse durations for which the
packet ends up in such a stage of its rapid oscillations and of
its slow drift that it is located at the nucleus and has zero
velocity. Second, a maximum, located independently of the pulse
duration, is due to a maximum lifetime of the KH bound state. For
such fields that more than one KH state exist and are
significantly populated, one additionally observes a third kind of
oscillations of the wave packet (in addition to the oscillations
with the optical frequency and the slow drift), which are due to
the population oscillating between the KH eigenstates.
\\

\acknowledgments{This work was supported in part by the Nicolaus
Copernicus University grant No. 442-F}



\begin{thebibliography}{99}
\bibitem{fedorov}
M.V. Fedorov, {\em Atomic and Free Electrons In a Strong Light
Field}, World Scientific, Singapore, 1997.
\bibitem{gavrila}
J.H. Eberly, R. Grobe, C.K. Law, Q. Su, in \emph{Atoms in Intense
Laser Fields}, edited by M. Gavrila (Academic Press, Boston,
1992), p. 301
\bibitem{hu}
S.X. Hu, L.A. Collins, Phys. Rev. A, 70, 013407 (2004)
\bibitem{mrz-stab}
J. Matulewski, A. Raczy\'nski and J. Zaremba, Phys. Rev. A, 61,
043402 (2000)
\bibitem{mrz-mg1}
J. Matulewski, A. Raczy\'nski and J. Zaremba, Phys. Rev. A, 68,
013408 (2003)
\bibitem{mrz-mg2}
J. Matulewski, A. Raczy\'nski and J. Zaremba, Phys. Rev. A, 68,
045401 (2003)
\bibitem{vps}
E.A. Volkova, A.M. Popov, O.V. Smirnova, Zh. Eksp. Teor. Fiz.,
106, 1360 (1994)
\bibitem{es}
J.H. Eberly, Q. Su, Phys. Rev. A, 44, 5997 (1991)
\bibitem{nrcpp}
W.H. Press, S.A. Teukolsky, W.T. Vetterling, B.P. Flannery, {\em
Numerical Recipes in C++. Second Edition}, Cambridge 2002.
\end{thebibliography}
\end{document}